# High-temperature magnetic anomaly via suppression of antisite disorder through synthesis route modification in a Kitaev candidate $Cu_2IrO_3$

Yuya Haraguchi[1,†], Daisuke Nishio-Hamane[2], Akira Matsuo[2], Koichi Kindo[2], and Hiroko Aruga Katori[1]
[1]Department of Applied Physics and Chemical Engineering, Tokyo University of Agriculture and Technology, Koganei, Tokyo 184-8588, Japan
[2]The Institute for Solid State Physics, The University of Tokyo, Kashiwa, Chiba 277-8581, Japan
†Corresponding author: chiyuya3@go.tuat.ac.jp

By incorporating inert KCl into the $Na_2IrO_3$ + 2CuCl → $Cu_2IrO_3$ + 2NaCl topochemical reaction, we significantly reduced the synthesis temperature of $Cu_2IrO_3$ from the 350°C reported in previous studies to 170°C. This adjustment decreased the Cu/Ir antisite disorder concentration in $Cu_2IrO_3$ from ~19% to ~5%. Furthermore, magnetic susceptibility measurements of the present $Cu_2IrO_3$ sample revealed a weak ferromagnetic-like anomaly with hysteresis at a magnetic transition temperature of ~70 K. Our research indicates that the spin-disordered ground state reported in chemically disordered $Cu_2IrO_3$ is an extrinsic phenomenon, rather than an intrinsic one, underscoring the pivotal role of synthetic chemistry in understanding the application of Kitaev model to realistic materials.

## I. Introduction

In quantum spin liquids (QSL), strong quantum fluctuations suppress magnetic ordering, resulting in a quantum state in which the spins are strongly entangled [1-3]. As a result, novel quasiparticles emerge that are different from conventional magnons in their magnetic excitations [4-6]. A spin model recognized by Ising-like bond-dependent interactions on a tricoordinated lattice (Kitaev model) is the most promising spin model that realizes a highly entangled QSL as a ground state [7]. The elementary excitations in Kitaev QSL are described by itinerant Majorana fermions and localized $Z_2$ fluxes emergent from the fractionalization of quantum spins [7]. Furthermore, the Kitaev model is materialized by the superexchange interactions between spin-orbital-coupled $J_{eff}$ = 1/2 electrons in $4d$/$5d$ transition metal ions with a low-spin $d^5$ electron configuration through the Jackeli-Khaliullin mechanism [8]. In this research trend, signatures of Kitaev QSL have been captured even in realistic materials [9-11], and then the materials developed to realize the Kitaev QSL ground state have been vigorously promoted worldwide [12-15].

One of the problems in realistic Kitaev materials is that the chemical disorders obscure the actual ground state [16-19]. Chemical disorders such as antisite disorder and stacking defects create a spatially fluctuating charge environment around magnetic ions, resulting in randomized pseudodipolar interactions. This randomness leads to a disordered state that mimics disordered states seemingly indistinguishable from quantum spin liquids in the form of superposition domains of short-range order [20]. In α-$Li_2IrO_3$, as an "old-school" Kitaev material, non-negligible antisite disorder on the honeycomb layer has been confirmed [21]. The disorder is controlled to some extent by the synthesis temperature. The magnetic order is observed as a cusp in the temperature dependence of the magnetic susceptibility at 15 K for the sample with minor antisite disorder. In contrast, the magnetic cusp is obscured for the sample with more significant antisite disorders.

Topochemical reactions have emerged as a promising strategy for creating novel Kitaev materials [10,11,14,22-25]. This method involves substituting the alkali metal ions in old-school Kitaev materials with a range of different isovalent/aliovalent cations, all under mild temperature conditions, which broadens the material search space for potential Kitaev physics. However, it simultaneously introduces a dilemma: while expanding the variety of materials, it also intensifies significant chemical disorders, including antisite disorder [14]. Especially in $Cu_2IrO_3$, synthesized via an ion exchange reaction of $Na_2IrO_3$ and CuCl at 350°C for 16 h, crystal structure analysis has clarified the presence of ~19% Cu and Ir antisite disorder [17, 23]. Furthermore, observing a spin-glass-like anomaly at 2.7 K also suggests the antisite disorder in $Cu_2IrO_3$. In addition, partial charge disproportionation has been observed as a chemical disorder specific to $Cu_2IrO_3$: about 10% of $Cu^+$ is oxidized to $Cu^{2+}$, and some $Ir^{4+}$ is reduced to $Ir^{3+}$ as charge compensation, as revealed by X-ray absorption spectroscopy experiments [26]. These facts demonstrate that the reaction temperature of 350°C is sufficient to produce these chemical disorders.

Meanwhile, attempts to observe the signatures of Kitaev

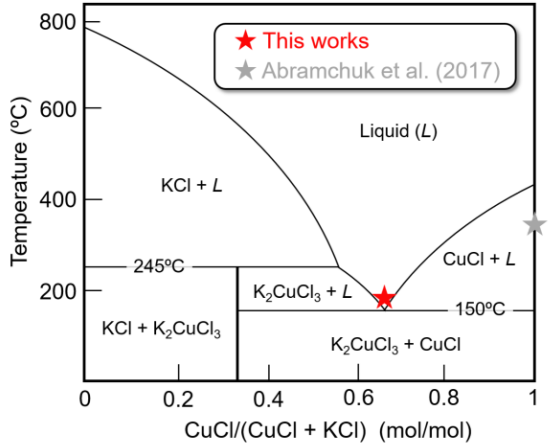

**Fig. 1** Phase diagram of CuCl-KCl system reconstructed from the previous report [29]. The gray and red stars indicate the temperature-composition synthesis conditions used in the topochemical reaction from $Na_2IrO_3$ to $Cu_2IrO_3$ in previous reports and in our work, respectively.

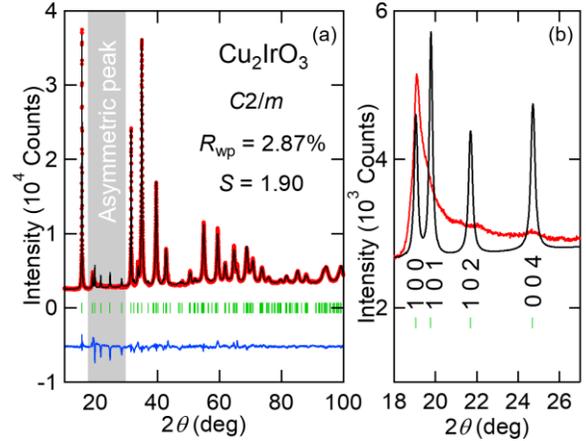

**Fig. 2** (a) The powder X-ray diffraction patterns of $Cu_2IrO_3$ synthesized at 170°C, showcasing observed intensities (in red), calculated intensities (in black), and their discrepancies (in blue). Vertical bars denote the Bragg reflection positions. The shaded region, where is excluded from the fitting range, emphasizes the characteristic asymmetric broadening of peak shapes, indicative of stacking faults. (b) The enlarged view of the asymmetric peak.

QSL in $Cu_2IrO_3$ continue, even considering the effects of the chemical disorder [27, 28]. However, the "smoking gun" for the Kitaev QSL feasibility in $Cu_2IrO_3$ has not yet been found in the present stage. Therefore, a different approach will be needed to solve these problems. Under this circumstance, we attempted to suppress the chemical disorder by seeking gentler synthetic conditions.

Our study aimed to reduce the reaction temperature by employing CuCl-KCl eutectic salts as the ion exchanger for the topochemical reaction instead of using CuCl alone. The pseudobinary phase diagram of CuCl and KCl, as reconstructed from the previous report [29], is depicted in Figure 1. The melting points of CuCl and KCl stand at 426°C and 770°C, respectively, while their eutectic mixture melts at approximately 150°C [29]. Through this approach, we successfully decreased the synthesis temperature of $Cu_2IrO_3$ from 350°C to 170°C. This alternative approach significantly decreased the Cu/Ir antisite disorder within the honeycomb layer, reducing it from ~19% to ~5%. Moreover, magnetic susceptibility measurements of the resulting sample indicated a weak ferromagnetic-like anomaly around 70 K, which contrasts with the low-temperature spin-glass behavior observed at 4 K in previously reported $Cu_2IrO_3$ samples. Therefore, our findings highlight the critical importance of addressing chemical disorders in Kitaev materials, as neglecting this issue can lead to incorrect assumptions regarding the ground state predicted by the Kitaev model.

## II. Experimental Methods

A precursor material, $Na_2IrO_3$, was prepared using the conventional solid-state reaction method following the previous report [30]: $Na_2CO_3$ and Ir in the ratio of 1.05:1 were intimately mixed, and the pelletized mixture placed on the corundum crucible was heated at 825°C for 48 h in air.

The initially reported synthesis of $Cu_2IrO_3$ was based on the topochemical reaction between precursor $Na_2IrO_3$ and CuCl.

This study explored more mild synthetic conditions using CuCl-KCl eutectic salts instead of CuCl. Using mixed salts is expected to increase the ionic diffusion coefficient and promote ion exchange reactions at lower temperatures. The phase diagram of the CuCl-KCl system [29] shows that the melting point drops to ~150°C at a eutectic point of CuCl/(CuCl+KCl) ~ 2/3 [see Fig. 1]. This precursor was ground well with a significant excess of CuCl-KCl mixture salt with the eutectic point ratio in an Ar-filled glove box, sealed in an evacuated Pyrex ampoule, and reacted at 170°C for 100 h. In addition, the ion exchange reaction is hardly promoted at temperatures just above and below the nominal eutectic point (150±5°C). Thus, the reaction temperature is maintained slightly above the nominal eutectic point to expedite the reaction through complete melting. The ion-exchange reaction is expressed as

$$Na_2IrO_3 + 2CuCl \rightarrow Cu_2IrO_3 + 2NaCl. \qquad (1)$$

After the reaction, the mixture was completely melted in the bottom of the Pyrex ampoule. The unreacted CuCl, inert KCl, and the byproduct NaCl were removed by washing the sample repeatedly with aqueous ammonia. The product was characterized by powder x-ray diffraction (XRD) experiments in a diffractometer with Cu-Kα radiation, and chemical analysis was conducted using a scanning electron microscope (JEOL IT- 100) equipped with an energy dispersive x-ray spectroscope (EDX) with 15 kV, 0.8 nA, 1-µm beam diameter). The cell parameters and crystal structure were refined by the Rietveld method using the Z-RIETVELD software [31].

The temperature dependence of the magnetization was measured under several magnetic fields using the magnetic property measurement system (MPMS; Quantum Design) at the Institute for Solid State Physics at the University of Tokyo. Magnetization curves up to ~60 T were measured by an induction method with a multilayer pulsed magnet at the International MegaGauss Science Laboratory of the Institute for Solid State Physics at the University of Tokyo. The temperature dependence of the heat capacity was measured using the conventional relaxation method in a physical property measurement system (PPMS; Quantum Design) at the Institute for Solid State Physics at the University of Tokyo.

**Table 1** The crystal parameters of $Cu_2IrO_3$ (space group $C2/m$) determined from powder X-ray diffraction experiments with the monoclinic unit cell lattice parameters $a = 5.39170(14)$ Å, $b = 9.3369(2)$ Å, $c = 5.97308(7)$ Å and $\beta = 107.823(2)°$. 'g' represents the atom occupancy parameter, and 'B' denotes the atomic displacement parameter.

| atom | site | g | x | y | z | B (Å²) |
|---|---|---|---|---|---|---|
| Ir1 | 4h | 0.9460(13) | 0 | 0.16843(9) | 1/2 | 0.117(12) |
| Ir2 | 2d | 0.108(3) | 0 | 0.5 | 1/2 | 0.40(4) |
| Cu1 | 4g | 1 | 0 | 0.3242(3) | 0 | 2.22(2) |
| Cu2 | 1a | 1 | 0 | 0 | 0 | 0.11(3) |
| Cu3 | 2d | 0.892(3) | 0 | 0.5 | 1/2 | 0.40(4) |
| Cu4 | 4h | 0.0540(13) | 0 | 0.16843(9) | 1/2 | 0.117(12) |
| O1 | 4i | 1 | 0.1158(19) | 0 | 0.3268(5) | 0.24(8) |
| O2 | 8j | 1 | 0.1066(12) | 0.3228(4) | 0.3338(4) | 0.13(5) |

### III. Results and Discussion

The powder XRD pattern of $Cu_2IrO_3$ synthesized at 170ºC is shown in Fig. 2. Despite the lower synthesis temperature, no residual precursor and impurity have been observed. The characteristic asymmetric broadening of specific peaks observed within the 2θ range of 15-25°, as highlighted in the shaded area, and its enlarged view shown in Fig. 2(b), is consistent with findings in previously reported $Cu_2IrO_3$ sample [23]. This broadening, also identified in similar layered materials, is attributed to stacking faults along the monoclinic c-axis [32-34].

The crystal structure of $Na_2IrO_3$ was initially identified as $C2/c$ [35], but subsequent studies reclassified it as $C2/m$ [36,37]. Given the similarity in the monoclinic structures of $Na_2IrO_3$ and $Cu_2IrO_3$, we conducted structural analyses for $Cu_2IrO_3$ employing both $C2/c$ and $C2/m$ models. These analyses yielded lower confidence parameters for the $C2/m$ structure, with a reliable S-factor of ~ 2.4 for the $C2/c$ structure and ~ 1.9 for the $C2/m$ structure. The comprehensive results of the Rietveld analysis employing the $C2/c$ model are extensively examined in the Supplemental Materials [38], which bolster the argument that the $C2/m$ structural model rather than $C2/c$ is more plausible from multiple chemical bonding viewpoints.

The chemical composition determined by EDX is 1.98(7) for Cu and 1.01(4) for Ir. Moreover, Na ion is not detected at all. These facts indicate that $Cu_2IrO_3$ is completely prepared even under gentle temperature conditions of 170°C. Our Rietveld refinement converges well with the crystal structure shown in Figs. 3(a) and 3(c) with crystallographic parameters in Table

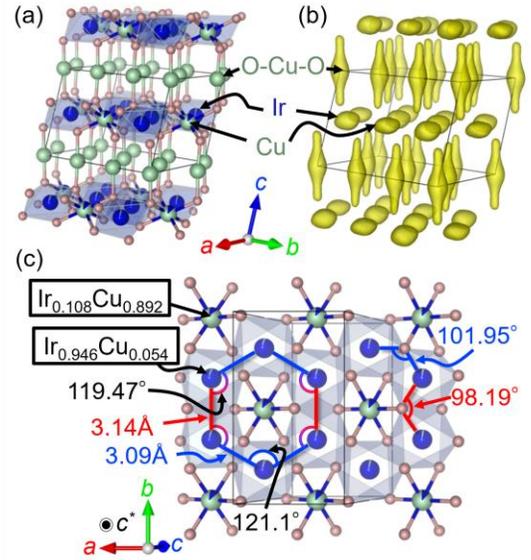

**Fig. 3** (a) Crystal structure of $Cu_2IrO_3$ ($T_{rxn} = 170°C$) viewed parallel the honeycomb layer. The light-green, blue, and light-pink spheres represent Cu, Ir, and O ions. The parallelepiped drawn by solid lines indicates the unit cell. (b) Three-dimensional nuclear density distributions of $Cu_2IrO_3$ calculated by the maximum entropy method (MEM) through the Z-RIETVELD software [31]. (c) The honeycomb layer structure of $Cu_2IrO_3$.

1. In our sample, the extent of site mixing between Cu and Ir ions within the honeycomb layer has been significantly reduced to approximately 5%, marking a substantial improvement over a previously reported figure of about 19% [23]. This advancement indicates that synthesizing the material under gentler temperature conditions has effectively promoted the exchange of constituent ions while preserving the crystal structural integrity of materials with minimal changes. In our $Cu_2IrO_3$ sample, synthesized at a reaction temperature of 170°C, the three Ir-O-Ir bond angles are measured at 98.19° and 101.95°, and the two Ir-Ir bond distances are determined to be 3.14 Å and 3.09 Å, respectively. These variations in the Ir-O-Ir superexchange pathways will likely influence the strength of interactions within the material.

The accuracy of the $C2/m$ structural model is further substantiated by the maximum entropy method (MEM) analysis. As illustrated in Fig. 3(b), the electron density contours effectively highlight the disparity between the electron densities associated with Ir and Cu atoms and the covalent characteristics of the Cu-O bond within the O-Cu-O dumbbell configuration. Additionally, it is noted in the Supplemental Materials that the O-Cu-O motifs are not observed in the MEM analysis outcomes when employing the $C2/c$ structural model [38], underscoring the distinctiveness of the $C2/m$ model in capturing these features.

In a comparative experiment, structural analysis of our synthesized samples, prepared using the same method previously reported, involving the reaction of $Na_2IrO_3$ and CuCl at 350°C for 16 hours, revealed approximately 19% antisite disorder between Ir and Cu sites within the honeycomb layer. This result is in good agreement with the previous report [23]. See the supplemental material for details on the structural analysis of our $T_{rxn}$ = 350°C sample [38]. These results indicated that the reaction temperature of 350°C is high-energy enough to generate some degree of Cu/Ir antisite disordering without thoroughly maintaining the Ir honeycomb framework. The bond valence sum (BVS) calculation for Ir ions from the refined structural parameters yields +4.09, consistent with the expected valence of +4. For comparison, the BVS calculation to the previously reported $Cu_2IrO_3$ sample ($T_{rxn}$ = 350°C) yields +3.72. This valence reduction is comparable to the results of X-ray absorption spectroscopy experiments [27].

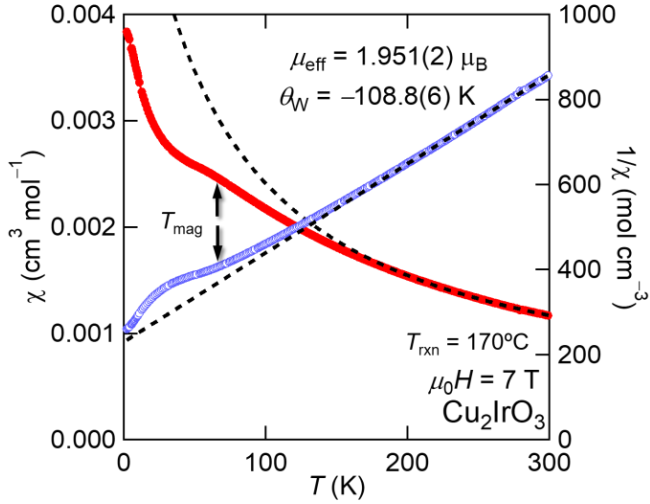

**Fig. 4** Temperature dependence of magnetic susceptibility $\chi$ and its inverse $1/\chi$ measured at 7 T of $Cu_2IrO_3$ prepared at 170°C. The dashed lines indicate the result of Curie-Weiss fitting.

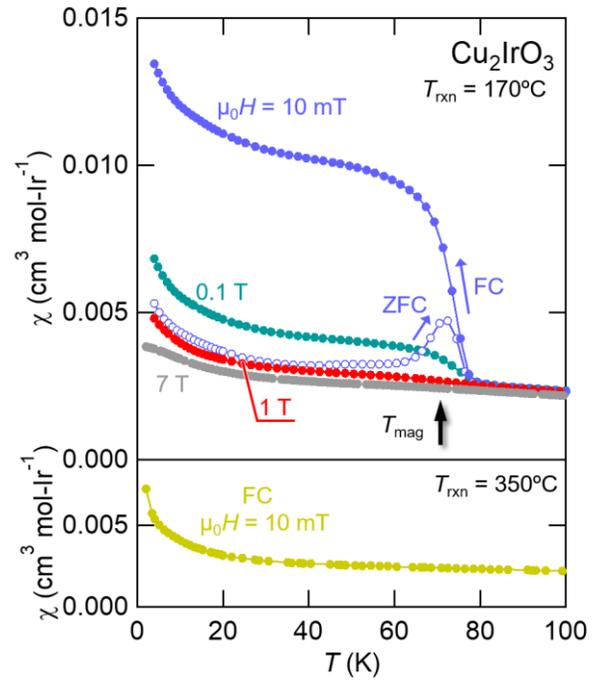

**Fig. 5** (Top panel) Temperature dependence of magnetic susceptibility $\chi$ in $Cu_2IrO_3$ prepared at 170°C measured at several magnetic fields. Ferromagnetic-like anomaly observed at $T_{mag}$ ~70 K. (Bottom panel) The $\chi$ data measured at 10 mT of $Cu_2IrO_3$ prepared at 350°C for comparison.

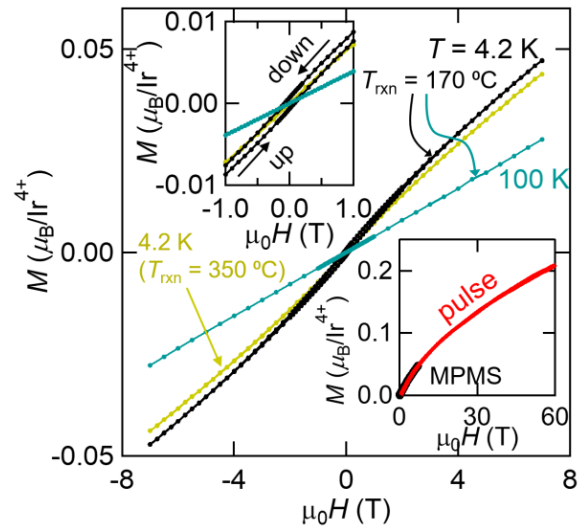

**Fig. 6** The isothermal magnetization $M$ curves in $Cu_2IrO_3$ ($T_{rxn}$ = 170°C) measured at 4.2 K (black) and 100 K (green) with the $M$ data in $Cu_2IrO_3$ prepared at 350°C for comparison. The upper inset shows the enlarged view of the magnetization curves near 0 T. The lower inset shows magnetization curve measured at 4.2 K under pulsed magnetic fields up to 60 T. The magnitudes are calibrated to the data measured under static fields up to 7 T.

Figure 4 shows the magnetic susceptibility $\chi$ and its inversed $1/\chi$ data for $Cu_2IrO_3$ ($T_{rxn}$ = 170°C) measured at 7 T. A linear relationship in $1/\chi$ versus $T$ in the high-temperature region indicates the presence of the local magnetic moment. A Curie-Weiss (CW) fitting of the inverse susceptibility at 200–300 K yields an effective magnetic moment $\mu_{eff}$ = 1.951(2) $\mu_B$ and Weiss temperature $\theta_W$ = −108.8(6) K. The negative $\theta_W$ value indicates that the interaction between $Ir^{4+}$ spins is predominantly antiferromagnetic. Furthermore, the $\mu_{eff}$ value suggests a $J_{eff}$ = 1/2 pseudospin system with a Landé g factor of $g$ = 2.25, which is enhanced by the spin-orbit interaction and is comparable to other iridates [35,39-41], including $Cu_2IrO_3$ ($T_{rxn}$ = 350°C) [23]. The $\chi$ data diverge from the Curie-Weiss (CW) fit at ~150 K and exhibit a weak anomaly at $T_{mag}$ ~ 70 K, indicative of either short-range or weak antiferromagnetic order. This characteristic was not observed in previously reported $Cu_2IrO_3$ samples ($T_{rxn}$ = 350°C) [23]. The emergence of this feature is likely attributed to the qualitative reduction of antisite disorder in the $Cu_2IrO_3$ sample, suggesting that improved synthesis conditions can significantly affect the magnetic properties.

The observed short-range ordering behavior near $T_{mag}$ in the

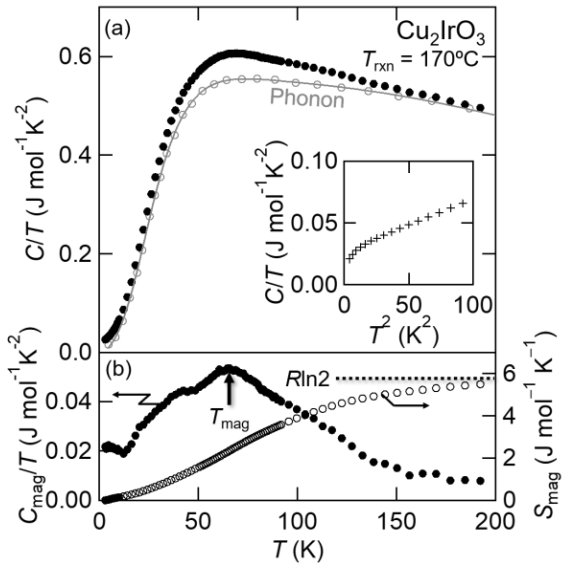

**Fig. 7** (a) The temperature dependence of the heat capacity divided by the temperature $C/T$ of $Cu_2IrO_3$. The inset shows the $C/T$ data at low temperatures as a function of $T^2$. Data is compared with the nonmagnetic analog $(Cu_{3/4}Li_{1/4})_2SnO_3$ [46] for estimation of the phonon contribution $C_{phonon}/T$ of $Cu_2IrO_3$. (b) The magnetic heat capacity $C_{mag}/T$ after the subtraction of $C_{phonon}/T$ and the magnetic entropy $S_{mag}$ for $Cu_2IrO_3$. The horizontal dashed line indicates the value of $S_{mag} = R\ln 2$, which is the total magnetic entropy derived from $J_{eff} = 1/2$ pseudospins.

$\chi$ data measured at 7 T closely resembles the short-range ordering identified in the promising Kitaev QSL candidate $H_3LiIr_2O_6$ [10]. To confirm the similarity in behavior, we conducted magnetization measurements on $Cu_2IrO_3$ ($T_{rxn}$ = 170°C) using various magnetic fields, the results of which are shown in Figure 5. Under a low magnetic field of 10 mT, the $\chi$ curve increases at ~70 K with a thermal hysteresis between the zero-field-cooled (ZFC) and field-cooled (FC) data. With an increasing magnetic field, the increase of the FC $\chi$ curve is suppressed and becomes almost indistinct at 7 T. A similar high-temperature magnetic anomaly was also observed in β-$Li_2IrO_3$ when subjected to a low magnetic field at 100 K [42]. This phenomenon can be coherently attributed to the reorientation of the principal axes of magnetization, a mechanism closely associated with the emergence of Kitaev-like correlations [42]. Currently, single crystals of $Cu_2IrO_3$ are unavailable, precluding the possibility of conducting detailed experiments to investigate changes in the easy axis. However, given that $Cu_2IrO_3$ is classified as a Kitaev magnetic material akin to β-$Li_2IrO_3$, it is plausible that similar magnetic anomalies could manifest in $Cu_2IrO_3$. The magnitude of a weak ferromagnetic moment does not differ for three differently prepared samples (see Supplemental Material [38]). On the other hand, as shown in the bottom panel, $Cu_2IrO_3$ ($T_{rxn}$ = 350°C) does not show such ferromagnetic enhancement, which is consistent with previous reports [23]. Hence, an emergent magnetic anomaly would be due to minimizing antisite disorder.

Figure 6 and its left inset show the isothermal magnetization $M$ curves at 100 and 4.2 K for $Cu_2IrO_3$ ($T_{rxn}$ = 170°C). The curve at 100 K is linear, while one at 4.2 K shows small spontaneous magnetization with hysteresis. This observation demonstrates the presence of a weak ferromagnetic moment accompanied by magnetic ordering. The spontaneous magnetic moment is about $6.8 \times 10^{-4}$ $\mu_B$ per $Ir^{4+}$ atoms. The origin of this tiny ferromagnetic moment will be discussed later. On the other hand, the absence of hysteresis in the $M$ data in $Cu_2IrO_3$ ($T_{rxn}$ = 350°C) at 4.2 K demonstrates that it is not a magnetically ordered state in contrast to $Cu_2IrO_3$ ($T_{rxn}$ = 170°C). The right inset of Fig. 6 shows the isothermal magnetization curve up to 60 T measured by an induction method with a multilayer pulsed magnet. In this field region, no metamagnetic anomalies involving the breakdown of magnetic order commonly seen in Kitaev magnets are observed [43,44].

Figure 7(a) presents the results of heat capacity measurements for $Cu_2IrO_3$ ($T_{rxn}$ = 170°C) and $(Cu_{3/4}Li_{1/4})_2SnO_3$. In these measurements, stannate is used to estimate the phonon contribution of the iridate, based on its well-known success as an effective nonmagnetic analog of Kitaev candidate iridates such as α-$Li_2IrO_3$ [45], $Na_2IrO_3$ [45], and $Ag_3LiIr_2O_6$ [22]. The $(Cu_{3/4}Li_{1/4})_2SnO_3$ powder samples used in these measurements were synthesized according to the method described in reference [46]. These measurements reveal that no distinct phase transition is detectable near $T_{mag}$. These observations resemble the magnetic anomalies observed at 100 K in β-$Li_2IrO_3$ [42]. Similarly, phenomena like those observed in spin glasses, marked by ferromagnetic moments that elude detection in heat capacity measurements, have been reported for pyrochlore iridates in studies [47,48]. Given that minor magnetic anomalies in pyrochlore iridates have been detected in heat capacity measurements using highly crystalline samples [49], it is plausible that magnetic anomalies at $T_{mag}$ of $Cu_2IrO_3$ could also become observable with enhanced crystallinity and reduced stacking faults. The inset of Figure 7 illustrates the $C/T$ data in the low-temperature region as a function of $T^2$, suggesting the presence of a finite γ-term behavior. This behavior indicates residual entropy, pointing to unresolved magnetic degrees of freedom within the system. Whether this observation stems from a Kitaev QSL state remains unclear in the present stage. In any case, clarifying the nature of the $T_{mag}$ anomaly is essential before any potential investigation into Kitaev QSL can proceed.

The magnetic contribution $C_{mag}/T$ was obtained by subtracting this phonon contribution from the experimental data. As shown in Fig. 7(b), the $C_{mag}/T$ curve exhibits a broad peak around $T_{mag}$ ~ 70 K, confirming that the magnetic anomaly in the χ-data is of bulk origin. Here, the accuracy of the estimation of the magnetic heat capacity of $Cu_2IrO_3$ should be carefully considered. Notably, the asymptotic value of $S_{mag}$, calculated by integrating the magnetic $C_{mag}/T$ with respect to $T$, coincides with the expected total magnetic entropy for $J_{eff} = 1/2$ pseudospin ($R\ln 2 = 5.76$ mol$^{-1}$K$^{-1}$) at high temperatures, as shown in Fig. 7(b). Additionally, the increase in $C_{mag}/T$ around 150 K is attributed to the development of short-range order, marking the point where the observed χ data and CW curve start to diverge. These observations demonstrate that $(Cu_{3/4}Li_{1/4})_2IrO_3$ serves as an effective non-magnetic analog for providing the lattice heat capacity contribution of $Cu_2IrO_3$, despite differences in phonon contributions due to the varying masses of Cu and Li, as well as Ir and Sn.

The $T_{mag}$ anomaly with a weak ferromagnetic moment probably originates from freezing a minor fraction of the spin degrees of freedom. This interpretation is consistent with the relatively low magnitude of the freezing/ordering moment, ascribed to the spontaneous magnetization of approximately 0.6% of the total magnetic moment of the $Ir^{4+}$ ion with $J_{eff} = 1/2$. According to the magnetic analog of the Clausius-Clapeyron equation, the change in magnetic entropy $\Delta S$ at the first-order phase transition is directly proportional to the increase in magnetization $\Delta M$ associated with spontaneous magnetization. Consequently, the latent heat associated with the transition at $T_{mag}$ is undetectably small. Therefore, it is reasonable to infer that anomalies at $T_{mag}$ would not result in detectable changes in entropy. To further understand the high-temperature magnetic anomaly in $Cu_2IrO_3$, whether it stems from the Kitaev model or spin-orbit entangled $J_{eff} = 1/2$ pseudospin physics, there is an urgent need to develop a method for preparing single-crystal samples. This development is under investigation, highlighting the importance of advancing sample preparation techniques to clarify the underlying mechanisms of these magnetic phenomena. Nevertheless, the $T_{mag}$ anomaly could serve as a crucial piece of the puzzle in understanding and resolving aspects of Kitaev physics.

From a wider perspective, optimized topochemical reactions under milder temperature conditions unveil the true physical properties previously obscured by chemical disorder. Therefore, strategies aimed at reducing the synthesis temperature could expose the genuine ground states of materials resembling spin liquids, where the impact of chemical disorder has traditionally been underestimated. Consequently, our findings suggest that overlooking chemical disorders could result in misleading conclusions.

## IV. Summary

Applying CuCl-KCl eutectic salts in the ion-exchange process from $Na_2IrO_3$ to $Cu_2IrO_3$ markedly lowered the reaction temperature from 350°C to 170°C. This adjustment in the synthesis approach led to the production of a $Cu_2IrO_3$ sample with significantly reduced antisite disorder between Ir and Cu from ~19% to ~5%, as verified through structural analysis. This sample demonstrated a magnetic anomaly near 70 K, featuring a slight ferromagnetic moment, marking a clear distinction from the spin glassy nonmagnetic ground state previously identified in $Cu_2IrO_3$ samples synthesized at 350°C. These findings underscore the critical role of synthesis conditions in defining the accurate structural and magnetic characteristics of complex oxides, suggesting that meticulous control over these parameters can unveil distinct physical properties.

## Acknowledgment

This work was supported by JST PRESTO Grant Number JPMJPR23Q8 (Creation of Future Materials by Expanding Materials Exploration Space) and JSPS KAKENHI Grant Numbers. JP23H04616 (Transformative Research Areas (A) "Supra-ceramics"), JP22K14002 (Young Scientific Research), and JP21K03441 (Scientific Research (C)). Part of this work was carried out by joint research in the Institute for Solid State Physics, the University of Tokyo (Project Numbers 202012-HMBXX-0021, 202112-HMBXX-0023, 202106-MCBXG-0065 and 202205-MCBXG-0063).